\documentclass{ws-ijmpd}
\usepackage[super,compress]{cite}
\begin{document}

\markboth{Sang Pyo Kim}
{Schwinger Effect, Hawking Radiation, and Unruh Effect}

%
\catchline{}{}{}{}{}
%

\title{SCHWINGER EFFECT, HAWKING RADIATION, AND UNRUH EFFECT}

\author{SANG PYO KIM}

\address{Department of Physics, Kunsan National University, Kunsan 54150, Korea \\
sangkim@kunsan.ac.kr}

\maketitle

\begin{history}
\received{Day Month Year}
\revised{Day Month Year}
\end{history}

\begin{abstract}
We revisit the Schwinger effect in de Sitter, anti-de Sitter spaces and charged black holes, and explore the interplay between quantum electrodynamics and the quantum gravity effect at one-loop level. We then advance a thermal interpretation of the Schwinger effect in curved spacetimes. Finally, we show that the Schwinger effect in a near-extremal black hole differs from Hawking radiation of charged particles in a non-extremal black hole and is factorized into those in an anti-de Sitter space and a Rindler space with the surface gravity for acceleration.
\end{abstract}


\ccode{PACS numbers: 04.62+v, 04.70.Dy, 12.20.-m}


\section{Introduction}\label{introduction}

Spontaneous creation of particles or charged pairs from an external gauge field or a curved spacetime is one of the most prominent phenomena in quantum field theory. A strong electric field produces pairs of charged particles and antiparticles, known as the Schwinger mechanism \cite{schwinger51}. The more remarkable phenomenon is the emission of all species of particles from black holes, known as Hawking radiation \cite{hawking74}. Under an influence of  strong backgrounds the vacuum may spontaneously breakdown due to quantum fluctuations and virtual pairs can be separated either by the energy of the fields or the causality of spacetimes. An accelerating detector measures a thermal spectrum of the Unruh temperature determined by the acceleration, known as the Unruh effect.\cite{unruh76} The spectrum and characteristics for these effects are summarized in Table \ref{ta1}.

Heisenberg and Euler found the one-loop effective action for an electron in a constant electromagnetic field\cite{heisenberg-euler36} and Schwinger introduced the proper-time integral method to express the effective action in scalar and spinor quantum electrodynamics (QED),\cite{schwinger51}  which is now known as the Heisenberg-Euler or Schwinger effective action. The most distinct feature of the Heisenberg-Euler or Schwinger action is poles of the proper-time representation of the action in an electric field. Thus, the one-loop effective action has not only the vacuum polarization (the real part) but also the vacuum persistence amplitude (twice the imaginary part). The vacuum persistence amplitude is a consequence of spontaneous production of charged pairs from the Dirac sea. Notice that the Schwinger effect is the particle-hole theory, in which virtual particles from the Dirac sea tunnel through a tilted potential barrier due to the electric field, and does not include the Coulomb attraction of pairs due to the homogeneity of produced pairs.
\begin{table}[ph]
\tbl{Comparison among the Schwinger mechanism, Unruh effect, and Hawking radiation.}
{\begin{tabular}{@{}cccc@{}} \toprule
  &{\bf Schwinger Mechanism} & {\bf Unruh Effect} & {\bf Hawking Radiation}\\ \colrule
Agent & Constant electric field & Accelerating detector & Charged RN black hole \\
Mean number & $N_{\rm S} = e^{-\frac{m}{2T_{\rm S}}}$ & $N_{\rm U} = \frac{1}{e^{\frac{\omega}{T_{\rm U}} } \mp 1}$  & $N_{\rm H} = \frac{\Gamma_{j \omega lm}}{e^{\frac{\omega-q_j \Phi}{T_{\rm H}}} \mp 1}$ \\
Temperature & $T_{\rm S} = \frac{1}{2 \pi} \frac{qE}{m}$ & $T_{\rm U} = \frac{a}{2 \pi}$  & $T_{\rm S} = \frac{1}{2 \pi} \frac{\sqrt{M^2 -Q^2}}{M + \sqrt{M^2 - Q^2}}$ \\
 \botrule
\end{tabular} \label{ta1}}
\end{table}

In this paper we recapitulate the Schwinger effect in curved spacetimes, such as a de Sitter $({\rm dS})$ space, an anti-de Sitter $({\rm AdS})$ space,
and an extremal or near-extremal Reissner-Nordstr\"{o}m (RN) black hole. One motivation for studying the Schwinger effect in ${\rm (A)dS}$ is to unveil the interplay between the Maxwell theory as a ${\rm U}(1)$ gauge and the quantum gravity effect at one-loop level. Another motivation to study QED in ${\rm (A)dS}$ is the near-horizon geometry of a near-extremal black hole\cite{CKLSW} and the scalar $S$-wave in the Nariai-geometry of a rotating black hole,\cite{Anninos-Hartman} which are summarized in Table \ref{ta2}. Further, it would be interesting to investigate whether charged black holes may have the Schwinger effect different from Hawking radiation.

\begin{table}[ph]
\tbl{QED phenomena in the near-horizon geometry of 4D black holes.}
{\begin{tabular}{@{}ccc@{}} \toprule
  {\bf Black Hole} & {\bf Region of Parameters} & {\bf Near-Horizon Geometry}\\ \colrule
RN black hole & Near-extremality $M \sim Q$ & QED in ${\rm AdS}_2 \times S^2$ \\
RN black hole & Non-extremality $M \gg Q$ & QED in ${\rm Rindler}_2 \times S^2$ \\
Rotating black hole in dS & scalar S-wave  & QED in ${\rm dS}_2$ \\
 \botrule
\end{tabular} \label{ta2}}
\end{table}

We also provide a thermal interpretation of the Schwinger effect in ${\rm (A)dS}_2$, which has recently been introduced by Cai and Kim\cite{cai-kim14,kim15c} and is a QED analog of the Unruh effect in ${\rm (A)dS}_2$.\cite{NPT96,deser-levin97} The Schwinger effect from an extremal RN black hole has the same spectrum as in ${\rm AdS}_2$ since the near-horizon geometry of the extremal black hole is ${\rm AdS}_2 \times S^2$ as shown in Table \ref{ta2}. The Schwinger formula from the extremal black hole is similarly given a thermal interpretation.\cite{KLY15,kim15a} Interestingly, the Schwinger effect from a near-extremal black hole is factorized into the Schwinger formula in ${\rm AdS}_2$ and that in ${\rm Rinder}_2$ with the acceleration of the surface gravity due to small non-extremality of black hole. We find the Schwinger formula in ${\rm dS}$ in any dimension. A passing remark is that the holographic Schwinger effect is the particle picture of charged pairs including the Coulomb attraction of the pairs.\cite{FNPT}

\section{QED Action in ${\rm dS}_2$}\label{sec3}

We consider the planar coordinates for a $(d+1)$-dimensional dS space, ${\rm dS}_{d+1}$,
\begin{eqnarray}
ds^2 &=& - dt^2 + e^{2Ht} d{\bf x}^2. \label{ds met}
\end{eqnarray}
The electromagnetic field in a curved spacetime measured in a local frame $\theta^{\alpha} = (dt, e^{Ht} dx^1, \cdots, e^{Ht} dx^{d})$ is given by the two-form tensor ${\bf F} = F_{\alpha \beta} \theta^{\alpha} \wedge \theta^{\beta}$. We assume a constant electric field along the $x^1$-direction measured by a co-moving observer with the $(d+1)$-velocity $u_{\alpha} = (1, 0, \cdots, 0)$. Then, ${\bf F}_{01} = - E = - {\bf F}_{10}$ in the local frame, ${\bf A} = - [E (e^{Ht} -1)/H] dx^1$ and ${\bf F} = d {\bf A}$ in the metric (\ref{ds met}). Thus, the vector potential is given by $A_1 = - E (e^{Ht} -1)/H$, which has the Minkowski limit $H = 0$.

First, in ${\rm dS}_2$ the Schwinger formula (mean number) for charged spinless scalars is given by the dimensionless instanton action\cite{garriga94,kim-page08}
\begin{eqnarray}
N_{\rm S} = e^{- S_{\rm dS}}, \quad  S_{\rm dS} = 2 \pi \Bigl( \sqrt{\bigl(\frac{qE}{H^2} \bigr)^2  + \bigl(\frac{m}{H} \bigr)^2 - \bigl(\frac{1}{2} \bigr)^2} - \frac{qE}{H^2} \Bigr). \label{ds sch}
\end{eqnarray}
The Schwinger formula (\ref{ds sch}) can be interpreted in terms of the effective temperature introduced by Cai and Kim\cite{cai-kim14}
\begin{eqnarray}
N_{\rm S} = e^{- \frac{\bar{m}}{T_{\rm CK}}}, \quad T_{\rm CK} = \sqrt{T_{\rm U}^2 + T_{\rm GH}^2} + T_{\rm U}, \label{CK ds tem}
\end{eqnarray}
where $T_{\rm U}$ is the Unruh temperature for accelerating charge, $T_{\rm GH}$ is the Gibbons-Hawking temperature,\cite{gibbons-hawking} and $\bar{m}$ is the effective mass in ${\rm dS}_2$, which are respectively
\begin{eqnarray}
T_{\rm U} = \frac{qE/\bar{m}}{2 \pi}, \quad T_{\rm GH} = \frac{H}{2 \pi}, \quad \bar{m} = \sqrt{m^2 - \bigl(\frac{H}{2} \bigr)^2}.
\end{eqnarray}
It is interesting to compare the effective temperature (\ref{CK ds tem}) with the effective temperature for an accelerating observer in ${\rm dS}_2$\cite{NPT96,deser-levin97}
\begin{eqnarray}
T_{\rm NPT-DL} = \sqrt{T_{\rm U}^2 + T_{\rm GH}^2} , \quad T_{\rm U} = \frac{a}{2 \pi}.
\end{eqnarray}

By solving the field equation and using the Bogoliubov transformation, the pair-production rate is found\cite{cai-kim14,kim15a,kim15c,haouat-chekireb15}
\begin{eqnarray}
\frac{d^2 N_{\rm dS}}{dt dx} = \frac{(2 \vert \sigma \vert +1) H^2 S_{\mu}}{2 (2 \pi)^2} \Bigl( \frac{e^{-(S_{\mu} - S_{\lambda})} \pm e^{-2 S_{\mu}}}{1 - e^{-2 S_{\mu}}} \Bigr), \label{ds2 sch}
\end{eqnarray}
where the upper (lower) sign is for scalars (fermions) and the dimensionless instanton actions are
\begin{eqnarray}
S_{\mu} = 2 \pi \sqrt{\bigl(\frac{qE}{H^2} \bigr)^2  + \bigl(\frac{m}{H} \bigr)^2 - \bigl[ \bigl(\frac{1}{2} \bigr)^2 \bigr]}, \quad S_{\lambda} = 2 \pi \frac{qE}{H^2}. \label{ds2 act}
\end{eqnarray}
The prefactor $H^2 S_{\mu}/2 (2 \pi)^2$ is the density of states and the square bracket in Eq. (\ref{ds2 act}) is present only for scalars but vanishes for fermions. Noting $S_{\rm dS} = S_{\mu} - S_{\lambda}$, the leading term of Eq. (\ref{ds2 sch}) is the Schwinger formula (\ref{ds sch}) from the instanton action. Without the density of states, the quantity in Eq. (\ref{ds2 sch}) is the mean number of created pairs.
In the in-out formalism, the vacuum persistence amplitude (integrated action density) is related to the mean number of pairs as
\begin{eqnarray}
2 {\rm Im} ({\rm W}_{\rm dS}) = \pm \bigl[\ln \bigl(1 \pm e^{-(S_{\mu} - S_{\lambda})} \bigr) - \ln \bigl(1- e^{-2 S_{\mu}} \bigr) \bigr], \label{vac per}
\end{eqnarray}
which is the pressure from quantum gas. The first logarithm in Eq. (\ref{vac per}) is the standard QED action with the mean number $N =  e^{-\bar{m}/T_{\rm CK}}$ while the second one is a correction due to a charged vacuum in ${\rm dS}$ and has the character of spinor QED regardless of spins.

Employing the gamma-function regularization in Ref.~\refcite{KLY08}, the QED action density for scalars is\cite{cai-kim14,kim15c}
\begin{eqnarray}
{\cal L}_{\rm dS}^{\rm sc} &=& \frac{H^2 S_{\mu}}{2 (2 \pi)^2} {\cal P} \int_{0}^{\infty} \frac{ds}{s} \Bigl[e^{-(S_{\mu} - S_{\lambda})s/2 \pi} \Bigl(\frac{1}{\sin(s/2)} - \bigl(\frac{2}{s} + \frac{s}{12} \bigr)  \Bigr) \nonumber\\&& - e^{- S_{\mu} s/\pi} \Bigl(\frac{\cos(s/2)}{\sin(s/2)} - \bigl(\frac{2}{s} - \frac{s}{6} \bigr)  \Bigr)  \Bigr].
\end{eqnarray}
Similarly, we find the QED action density for fermions
\begin{eqnarray}
{\cal L}_{\rm dS}^{\rm sp} = - \frac{H^2 S_{\mu}}{(2 \pi)^2} {\cal P} \int_{0}^{\infty} \frac{ds}{s} \Bigl(e^{-(S_{\mu} - S_{\lambda})s/2\pi} - e^{- S_{\mu}s/\pi} \Bigr)\Bigl(\frac{\cos(s/2)}{\sin(s/2)} - \bigl(\frac{2}{s} - \frac{s}{6} \bigr) \Bigr).
\end{eqnarray}
Here, ${\cal P}$ denotes the principal value and the subtracted terms renormalize the vacuum energy and the charge.
The bosonic and fermionic current density is given by
\begin{eqnarray}
J_{\rm dS} = 2q \frac{d^2 N_{\rm dS}}{dt dx}. \label{cur ds}
\end{eqnarray}
Note that the current density comes from the second quantized field theory in curved spacetime, which is equivalent to $(2~{\rm charge}) \times ({\rm density~ of~ states}) \times ({\rm mean~ number})$. In the limit of $mH \leq qE \ll H^2$, i.e., $S_{\mu} \approx (qE/H^2)$ and $S_{\rm dS} \approx \pi (mH/qE)^2$, the current density accumulated over the Hubble time $1/H$ increases as $(H^3/E)\times (2 \pi m^2)$ and exhibits the infrared hyperconductivity, which has been observed in Ref.~\refcite{FGKSSTV14}.

\section{QED Action in ${\rm dS}_{d+1}$}\label{QED dS2}

The Schwinger effect in ${\rm dS}_4$ has been studied in Refs.~\refcite{kobayashi-afshordi,stahl-strobel}. We now study the Schwinger effect in $(d+1)$ dimensions, in which the field equation for charged scalars takes the form
\begin{eqnarray}
 \ddot{\phi}_{\bf k} + dH \dot{\phi}_{\bf k} +\Bigl(m^2  + e^{-2 Ht}\bigl(\bar{k}_1 + \frac{qE}{H} e^{Ht} \bigr)^2 + {\bf k}_{\perp}^2 e^{-2Ht} \Bigr) \phi_{\bf k} = 0,
\end{eqnarray}
where $\bar{k}_1 = k_1 - qE/H$ and ${\bf k}_{\perp}$ is the $(d-1)$-dimensional momentum transverse to the electric field. Then, the positive and negative frequency solutions for the in-vacuum ($t = - \infty$) are given by the Whittaker function as
\begin{eqnarray}
\phi^{(+)}_{{\rm in}, {\bf k}} (t) &=& \frac{e^{\pi \kappa/2}}{\sqrt{2 \bar{k}}} W_{-i \kappa, i \gamma_{\rm dS}} (2 e^{-i \pi/2} \frac{\bar{k}}{H}  e^{-Ht}), \nonumber\\
\phi^{(-)}_{{\rm in}, {\bf k}} (t) &=& \frac{e^{\pi \kappa/2}}{\sqrt{2 \bar{k}}} W_{i \kappa, i \gamma_{\rm dS}} (2 e^{i \pi/2} \frac{\bar{k}}{H} e^{-Ht}), \label{dsD in-sol}
\end{eqnarray}
where
\begin{eqnarray}
\kappa = - \frac{qE}{H^2} \frac{\bar{k}_{1}}{\bar{k}}, \quad \bar{k} = \sqrt{\bar{k}_1^2 + {\bf k}_{\perp}^2}, \quad  \gamma_{\rm dS}  = \sqrt{\bigl(\frac{qE}{H^2} \bigr)^2  + \bigl(\frac{m}{H} \bigr)^2 - \bigl(\frac{d}{2} \bigr)^2}.
\end{eqnarray}
The positive and negative frequency solutions for the out-vacuum ($t = \infty$) are given by the confluent hypergeometric function as
\begin{eqnarray}
\phi^{(+)}_{{\rm out}, k} (t) &=& \frac{e^{- i \pi/4} e^{\pi \gamma_{\rm dS}/2}}{\sqrt{4 \bar{k} \gamma_{\rm dS}}} M_{i \kappa, i \gamma_{\rm dS}} (2 e^{i \pi/2} \frac{\bar{k}}{H} e^{-Ht}), \nonumber\\
\phi^{(-)}_{{\rm out}, k} (t) &=& \frac{e^{- i \pi/4} e^{- \pi \gamma_{\rm dS}/2}}{\sqrt{4 \bar{k} \gamma_{\rm dS}}} M_{i \kappa, - i \gamma_{\rm dS}} (2 e^{i \pi/2} \frac{\bar{k}}{H} e^{-Ht}). \label{dsD out-sol}
\end{eqnarray}
The Schwinger effect in a constant electric field in ${\rm dS}_{d+1}$ should be independent of $t$ and $x_1$ due the symmetry of the spacetime and the field. In fact, the integration of the longitudinal momentum $k_1$ of the out-vacuum solution $\phi^{(+)}_{{\rm out}, k} (t)$ gives the density of states $H^2 \gamma_{\rm dS}/(2 \pi)$ independent of dimensions.\cite{kim15c} Therefore, the pair-production rate per spacetime volume is
\begin{eqnarray}
\frac{d^{d+1}N_{\rm dS}}{dt dx^d} = \frac{(2 \vert \sigma \vert +1) H^2 S_{\mu}}{2 (2 \pi)^2} \int \frac{d^{d-1} {\bf k}_{\perp}}{(2 \pi)^{d-1}}
\Bigl( \frac{e^{-(S_{\mu} - S_{\lambda})} \pm e^{- 2 S_{\mu}}}{1-e^{- 2 S_{\mu}}} \Bigr), \label{D-dS}
\end{eqnarray}
where the upper (lower) sign is for charged scalars (fermions) and the dimensionless instanton actions are
\begin{eqnarray}
S_{\mu} = 2 \pi \sqrt{\bigl(\frac{qE}{H^2} \bigr)^2  + \bigl(\frac{m}{H} \bigr)^2 - \bigl[ \bigl(\frac{d}{2} \bigr)^2 \bigr]}, \quad S_{\lambda} = 2 \pi \frac{qE}{H^2} \Bigl( \frac{\frac{qE}{H}}{\sqrt{ \bigl( \frac{qE}{H} \bigr)^2 + {\bf k}_{\perp}^2 }} \Bigr). \label{D-dS ins}
\end{eqnarray}
Here, the square bracket in Eq. (\ref{D-dS ins}) vanishes for fermions. In the case of ${\rm dS}_2$, the transverse momentum vanishes and the integration becomes unity and the Schwinger formula (\ref{ds2 sch}) is recovered. Further, in the limit of the Minkowski spacetime ($H = 0$),  Eq. (\ref{D-dS}) recovers the Schwinger formula ($d=3$ in Table \ref{ta1})
\begin{eqnarray}
\frac{d^{d+1} N_{\rm Min}}{dt dx^d} = \frac{(2 \vert \sigma \vert +1) qE}{2 (2 \pi)} \int \frac{d^{d-1} {\bf k}_{\perp}}{(2 \pi)^{d-1}} e^{-\pi \frac{m^2 + {\bf k}_{\perp}^2}{qE}}. \label{D-Min}
\end{eqnarray}
Finally, in the limit of the zero field $(E=0)$, Eq. (\ref{D-dS}) reduces to the pure dS radiation for scalars (fermions) in the planar coordinates
\begin{eqnarray}
N_{\rm dS} = \frac{1}{e^{S_{\rm dS}}\mp 1}, \quad S_{\rm dS} = 2 \pi \sqrt{\bigl(\frac{m}{H} \bigr)^2 - \bigl[ \bigl(\frac{d}{2} \bigr)^2 \bigr]}.
\end{eqnarray}

\section{QED Action in ${\rm AdS}_2$}\label{QED ads2}

The two-dimensional ${\rm AdS}$ space has a constant curvature ${\cal R}_2 = -2K^2$, the planar coordinates and the Coulomb potential for a constant electric field
\begin{eqnarray}
ds^2 = - e^{2Kx} dt^2 + dx^2, \quad A_1 = - \frac{E}{K} (e^{Kx} -1).
\end{eqnarray}
The Schwinger formula for charged scalars is given by another dimensionless instanton action\cite{pioline-troost,kim-page08}
\begin{eqnarray}
N_{\rm AdS} = e^{-S_{\rm AdS}} , \quad     S_{\rm AdS} = 2 \pi \Bigl( \frac{qE}{K^2} -  \sqrt{\bigl(\frac{qE}{K^2} \bigr)^2  - \bigl(\frac{m}{K} \bigr)^2 - \bigl(\frac{1}{2} \bigr)^2} \Bigr), \label{ads sch}
\end{eqnarray}
which is an analytical continuation of the instanton action (\ref{ds sch}) in ${\rm dS}_2$.
Employing the effective temperature by Cai and Kim\cite{cai-kim14}, one may give a thermal interpretation for the Schwinger formula (\ref{ads sch}) as
\begin{eqnarray}
N_{\rm AdS} = e^{- \frac{\bar{m}}{T_{\rm CK}}}, \quad T_{\rm CK} = \sqrt{T_{\rm U}^2 + \frac{{\cal R}_2}{8 \pi^2}} + T_{\rm U}, \label{CK tem}
\end{eqnarray}
where the Unruh temperature for accelerating charge and the effective mass in ${\rm (A)dS}_2$ are
\begin{eqnarray}
T_{\rm U} = \frac{qE/\bar{m}}{2 \pi}, \quad \bar{m} = \sqrt{m^2 - \frac{{\cal R}_2}{8}}.
\end{eqnarray}
The binding nature of a pair in the ${\rm AdS}$ space increases the effective mass while the ${\rm dS}$ space intrinsically separates the pair and decreases the effective mass.

Similarly, one obtains the density of states and the pair-production rate in ${\rm AdS}_2$
\begin{eqnarray}
\frac{d^2 N_{\rm AdS}}{dt dx} = \frac{(2 \vert \sigma \vert +1) K^2 S_{\nu}}{2 (2 \pi)^2}\Bigl( \frac{e^{-(S_{\lambda} - S_{\nu})} - e^{- (S_{\lambda} + S_{\nu})}}{1 \pm e^{- (S_{\lambda} + S_{\nu})} } \Bigr), \label{ads2 sch}
\end{eqnarray}
where the upper (lower) sign is for charged scalars (fermions) and the dimensionless instanton actions are
\begin{eqnarray}
S_{\lambda} = 2 \pi \frac{qE}{K^2}, \quad S_{\nu} = 2 \pi \sqrt{\bigl(\frac{qE}{K^2} \bigr)^2  - \bigl(\frac{m}{K} \bigr)^2 - \bigl[ \bigl(\frac{1}{2} \bigr)^2 \bigr]}.
\end{eqnarray}
The square bracket vanishes for fermions. Note that pairs are produced when the Breitenlohner-Freedman stability bound $m^2 \geq (qE/K)^2$ is violated.
The QED action density is for scalars
\begin{eqnarray}
{\cal L}_{\rm AdS}^{\rm sc} = \frac{K^2 S_{\nu}}{2 (2 \pi)^2} {\cal P} \int_{0}^{\infty} \frac{ds}{s} e^{-S_{\lambda}s/2\pi} \cosh( S_{\nu}s/2 \pi) \Bigl(\frac{1}{\sin(s/2)} - \bigl(\frac{2}{s} + \frac{s}{12} \bigr)  \Bigr),
\end{eqnarray}
and for fermions
\begin{eqnarray}
{\cal L}_{\rm AdS}^{\rm sp} &=& - \frac{K^2 S_{\nu}}{(2 \pi)^2} {\cal P} \int_{0}^{\infty} \frac{ds}{s} \Bigl(e^{-(S_{\lambda} - S_{\nu})s/2\pi} - e^{- (S_{\lambda}+ S_{\nu})s/2\pi} \Bigr) \nonumber\\ && \times \Bigl(\frac{\cos(s/2)}{\sin(s/2)} - \bigl(\frac{2}{s} - \frac{s}{6} \bigr) \Bigr).
\end{eqnarray}
The bosonic and fermionic current density is given by
\begin{eqnarray}
J_{\rm AdS} = 2q \frac{d^2 N_{\rm AdS}}{dt dx}.
\end{eqnarray}

\section{Hawking Radiation from Non-Extremal Black Hole}\label{Non BH}

The RN black hole with the mass $M$ and charge $Q$ has the metric and the Coulomb potential
\begin{eqnarray}
ds^2 = - \Bigl(1 - \frac{2M}{r} + \frac{Q^2}{r^2} \Bigr)dt^2 + \frac{dr^2}{1 - \frac{2M}{r} + \frac{Q^2}{r^2}} + r^2 d\Omega_2^2, \quad \Phi = \frac{Q}{r}, \label{rn bh}
\end{eqnarray}
which has the outer and inner horizons
\begin{eqnarray}
r_{+} = M + \sqrt{M^2 - Q^2}, \quad r_{-} = M - \sqrt{M^2 - Q^2}.
\end{eqnarray}
As summarized in Table \ref{ta2}, the near-horizon geometry of the RN black hole (\ref{rn bh}) is approximately given by the Rindler space, ${\rm Rindler}_2 \times S^2$ when $M \gg Q$ while it is accurately approximated by the ${\rm AdS}$ space, ${\rm AdS}_2 \times S^2$ when $M \simeq Q$. Interestingly, in the tortoise coordinate $r_*$ for an extremal black hole $Q = M$, the radial motion of a charged scalar in the spherical harmonics $Y_{lm} (\theta, \varphi)$,\cite{kim14a}
\begin{eqnarray}
\Bigl[ \frac{d^2}{dr_*^2} - V_l (r(r_*)) \Bigr] \Phi_{l} = 0
\end{eqnarray}
has in the asymptotic region $(r \gg Q)$ the effective potential for the Klein-Gordon equation in the Coulomb potential with an angular momentum $\bar{l}$ and a charge $\bar{Q}$
\begin{eqnarray}
V_l (r) = m^2 + \frac{\bar{l}(\bar{l}+1)}{r^2} - \Bigl(\omega - \frac{q \bar{Q}}{r} \Bigr)^2 + {\cal O} \Bigl( \frac{1}{r^3} \Bigr), \label{eff pot}
\end{eqnarray}
where
\begin{eqnarray}
\bar{Q} = Q \Bigl(1 + \frac{m^2}{q \omega} \Bigr), \quad \bar{l}+ \frac{1}{2} = \sqrt{\Bigl( l+ \frac{1}{2} \Bigr)^2 + (qQ)^2 \Bigl( \frac{m^2}{q \omega} + 1 \Bigr)^2 - (qQ)^2}.
\end{eqnarray}
Thus, the Schwinger formula from the extremal black hole is approximately given by that from a supercritical point charge in the Minkowski spacetime.

In the tunneling picture for Hawking radiation, virtual pairs are created through the horizon. Thus, the near-horizon geometry of the black hole plays an important role in understanding the Schwinger effect or Hawking radiation. For a non-extremal black hole, the near-horizon geometry has the Rindler coordinates\cite{kim07}
\begin{eqnarray}
ds^2 = - F(\rho) d\tau^2 + \frac{d\rho^2}{G(\rho)} + r^2 (\rho) d \Omega_2^2,
\label{metric}
\end{eqnarray}
where $F \simeq (\kappa \rho)^2$ and $G \simeq 1$ near the horizon $\rho  \simeq 0$ with the surface gravity $\kappa = (r_+-r_-)/(2 r_+^2)$.
A physically intuitive way to understand the pair production from the background field is to investigate the poles of the Hamilton-Jacobi equation for a charged particle. The Hamilton-Jacobi action for each spherical harmonics
\begin{eqnarray}
S_{l} = \int \frac{d \rho}{\sqrt{FG}} \sqrt{(\omega - q \Phi)^2 - F \Bigl(\frac{(l+\frac{1}{2})^2}{r^2 (\rho) }+  m^2 \Bigr)} \label{rn act}
\end{eqnarray}
can give the particle production as the decay rate of the vacuum. In the phase-integral method, the mean number is given by the leading term\cite{kim07,kim-page07,kim13}
\begin{eqnarray}
{\cal N}_l = e^{i  {\cal S}_{\Gamma_l}},
\end{eqnarray}
where ${\cal S}_{\Gamma_l} = \oint S_l $ is the action evaluated along a contour $\Gamma_l$ in the complex plane of $\rho$.
The simple pole at the outer horizon $\rho = 0~(r = r_+)$ recovers Hawking radiation of charged particles
\begin{eqnarray}
{\cal N}_l = e^{- \frac{\omega - q \Phi (r_+) }{T_{\rm H}}}, \label{hawking rad}
\end{eqnarray}
where the Hawking temperature and the electrostatic potential on the horizon are
\begin{eqnarray}
T_{\rm H} = \frac{r_+ - r_-}{4 \pi r_+^2}, \quad \Phi (r_+) =  \frac{Q}{r_+}.
\end{eqnarray}

\section{Schwinger Effect in Near-Extremal Black Hole}\label{NE BH}

As shown in Table \ref{ta2}, the near-horizon geometry of a near-extremal black hole has ${\rm AdS}_2 \times S^2$ with the metric\cite{CKLSW}
\begin{eqnarray}
ds^2 = - \frac{\rho^2 - B^2}{Q^2} d \tau^2 + \frac{Q^2}{\rho^2 - B^2} d \rho^2 + Q^2 d\Omega_2^2, \label{near geom}
\end{eqnarray}
where coordinates are stretched by smallness parameter $\epsilon$ as
\begin{eqnarray}
r - Q = \epsilon Q, \quad t = \frac{\tau}{Q}, \quad M - Q = \frac{(\epsilon B)^2}{2Q}. \label{bh par}
\end{eqnarray}
The Hawking radiation for the parametrization (\ref{bh par}) is given by
\begin{eqnarray}
N_{\rm H} = \frac{1}{e^{\frac{\omega - q \Phi (B)}{T_{\rm H} }} - 1}, \quad T_{\rm H} = \frac{\sqrt{1/2 + M/2Q}}{4 \pi (M + \epsilon B \sqrt{1/2+M/2Q})} \epsilon B.
\end{eqnarray}
Except for particles with energy close to the chemical potential, i.e, $\omega - q \Phi (B) \propto \epsilon B$, Hawking radiation is exponentially suppressed to zero. By taking the limit $\epsilon = 0$ the near-horizon region can be stretched infinitely to ${\rm AdS}_2 \times S^2$. This procedure differs from the amplified near-horizon geometry of a non-extremal black hole as ${\rm Rindler}_2 \times S^2$. This means that the Schwinger effect from ${\rm AdS}_2 \times S^2$ is very accurate for the near-extremal black hole.

The action for the Hamilton-Jacobi equation for the radial motion after separating the spherical harmonics is given by
\begin{eqnarray}
S_{l} (\rho) = \int d \rho \sqrt{\frac{(q\rho - \omega Q)^2 Q^2}{(\rho^2 - B^2)^2} - \frac{m^2 Q^2 + (l+1/2)^2}{\rho^2 - B^2}}.
\end{eqnarray}
Note that there are two finite poles at $\rho = B$, the outer horizon, and $\rho = -B$, the inner horizon as well as an infinite pole at $\rho = \infty$.
Then, the contour integral gives the leading term for the Schwinger effect\cite{CKLSW}
\begin{eqnarray}
N_{\rm NBH} = e^{-(S_a - S_b)}, \quad S_a = 2 \pi qQ, \quad S_b = 2 \pi \sqrt{(q^2 - m^2) Q^2 -(l+1/2)^2},
\end{eqnarray}
where $S_a$ is the sum of residues at $\rho = \pm B$ and $S_b$ is the residue at $\rho = \infty$.

From the field equation, the exact Schwinger formula from the geometry (\ref{near geom}) is given by\cite{CKLSW}
\begin{eqnarray}
N_{\rm NBH} = \Bigl( \frac{e^{-(S_a - S_b)} - e^{-(S_a + S_b)}}{1 \pm e^{-(S_a + S_b)}} \Bigr) \times \Bigl( \frac{1 \mp e^{-(S_c - S_a)}}{1 + e^{-(S_c - S_b)}} \Bigr) \label{nbh mean}
\end{eqnarray}
where $S_c = 2 \pi (\omega Q^2/B)$. The mean number for the Schwinger effect has the thermal interpretation\cite{KLY15,kim15a}
\begin{eqnarray}
N_{\rm NBH} = \Bigl( \frac{e^{- \frac{\bar{m} }{T_{\rm CK} }} - e^{- \frac{\bar{m} }{\bar{T}_{\rm CK} }} }{1 \pm e^{- \frac{\bar{m} }{\bar{T}_{\rm CK} } } } \Bigr) \times \Bigl( \frac{1 \mp e^{- \frac{\omega - q \Phi}{T_{\rm H} }} }{1 + e^{-\frac{\omega - q \Phi}{T_{\rm H}} - \frac{\bar{m} }{T_{\rm CK} } } } \Bigr), \label{nbh ther}
\end{eqnarray}
where
\begin{eqnarray}
T_{\rm CK} = T_{\rm U} + \sqrt{T_{\rm U}^2 - \bigl(\frac{1}{2 \pi Q} \bigr)^2}, \quad \bar{T}_{\rm CK} = T_{\rm U} - \sqrt{T_{\rm U}^2 - \bigl(\frac{1}{2 \pi Q} \bigr)^2} \label{nbh tem}
\end{eqnarray}
with $T_{\rm U} = (qE_{\rm H}/\bar{m})/2 \pi$, the Unruh temperature for accelerating charge on the horizon. Note that $T_{\rm CK}$ and $\bar{T}_{\rm CK}$ are the temperature for the Schwinger effect in ${\rm AdS}_2$ in Sec. \ref{QED ads2}. The Schwinger effect from the near-extremal black hole is the product of the Schwinger effect in ${\rm AdS}_2$ and a correction due to the Hawking radiation from the non-extremality. It has been further observed that the Schwinger effect has the factorization\cite{KLY15,kim15a}
\begin{eqnarray}
N_{\rm NBH} = e^{\frac{\bar{m} }{T_{\rm CK} }} \times \underbrace{ \Bigl( \frac{e^{- \frac{\bar{m} }{T_{\rm CK} }} - e^{- \frac{\bar{m} }{\bar{T}_{\rm CK} }} }{1 \pm e^{- \frac{\bar{m} }{\bar{T}_{\rm CK} } } } \Bigr)}_{\rm Schwinger~effect~in~AdS_2}  \times \underbrace{\Bigl( \frac{ e^{- \frac{\bar{m} }{T_{\rm CK}}} (1 \mp e^{- \frac{\omega - q \Phi}{T_{\rm H} }}) }{1 + e^{-\frac{\omega - q \Phi}{T_{\rm H}} - \frac{\bar{m} }{T_{\rm CK} } } } \Bigr)}_{\rm Schwinger~effect~in~Rindler_2}. \label{nbh rin}
\end{eqnarray}
The reason for the Schwinger effect in ${\rm Rindler}_2$ is that the measure of the near-extremality is in fact a measure for the Rindler space. The Schwinger effect of scalar QED in the Rindler space has been studied in Ref.~\refcite{gabriel-spindel00}.

\section{Discussion and Conclusion}\label{discussion}

We have studied the spontaneous creation of charged pairs from a constant electric field in ${\rm (A)dS}$ and from a charged RN black hole, which is a nonperturbative quantum effect at one-loop level. The QED action in ${\rm (A)dS}$ exhibits the interplay of QED and the quantum gravity effect. The Schwinger formula has a thermal interpretation of the Unruh temperature for accelerating charge and the Gibbons-Hawking temperature for ${\rm dS}_2$ and the corresponding curvature for ${\rm AdS}_2$. We have found the Schwinger formula in ${\rm dS}$ space of any dimension, which has the correct limit of the Minkowski spacetime and the pure ${\rm dS}$ space. The near-horizon geometry of charged RN black hole is ${\rm AdS}_2 \times S^2$ for an extremal black hole while it is ${\rm Rindler}_2 \times S^2$ for a non-extremal black hole. Thus, the Schwinger effect for extremal black hole has the same thermal interpretation as in ${\rm AdS}_2$. The near-extremal black hole has, however, an additional Schwinger effect in ${\rm Rindler}_2$.

\section*{Acknowledgements}
The author would like to thank Rong-Gen Cai, Chiang-Mei Chen, W-Y. Pauchy Hwang, Hyun  Kyu Lee, Don N. Page, Christian Schubert, and Yongsung Yoon for useful discussions and Pisin Chen for warm hospitality during the second LeCosPA Symposium. The participation of LeCosPA Symposium was supported by Asia Pacific Center for Theoretical Physics (APCTP). This research was supported by Basic Science Research Program through the National Research Foundation of Korea(NRF) funded by the Ministry of Education (15B15770630).

\end{document}